\documentclass[11pt]{article}
\usepackage{vietnam,epsfig}

\def\Journal#1#2#3#4{{#1} {\bf #2}, #3 (#4)}

\def\NPA{{\em Nucl. Phys.} A}

\def\be{\begin{equation}}
\def\ee{\end{equation}}
\def\bea{\begin{eqnarray}}
\def\eea{\end{eqnarray}}

\begin{document}
\vspace*{4cm}
\title{CHALLENGE ON $^{48}$Ca ENRICHMENT \\
FOR CANDLES DOUBLE BETA DECAY EXPERIMENT}

\author{R. HAZAMA$^1$, Y. TATEWAKI$^2$, T. KISHIMOTO$^2$, K. MATSUOKA$^2$, \\
N. ENDO$^3$, K. KUME$^3$, Y. SHIBAHARA$^4$, and M. TANIMIZU$^5$}

\address{$^1$Department of Engineering, Hiroshima University, Higashi-Hiroshima 739-8527, Japan\\
$^2$Department of Physics, Osaka University, Toyonaka, Osaka, 560-0043, Japan\\
$^3$The Wakasa Wan Energy Research Center, Tsuruga, Fukui 914-0192, Japan\\
$^4$Department of Nuclear Engineering, Osaka University, Suita, Osaka, 565-0871, Japan\\
$^5$Institute for Research on Earth Evolution (IFREE), 
Japan Agency for Marine-Earth Science and Technology (JAMSTEC), Yokosuka, Kanagawa 237-0061, Japan}

\maketitle\abstracts{
Chemical isotope effects of calcium were studied by liquid-liquid extraction using 
a crown ether of dicyclohexano-18-crown-6 for the purpose of finding a 
cost-effective and efficient way of enrichment of $^{48}$Ca towards the 
study of the neutrinoless double beta decay ($\beta\beta$) of $^{48}$Ca.  
We evaluated each contribution ratio of the field shift effect and the hyperfine 
splitting shift effect to the mass effect of the calcium isotopes for the first time.  
The present preliminary result suggests the contribution of the field shift effect 
is small, especially for $^{40}$Ca$-^{48}$Ca case, compared with 
the case of Chromium trichloride-crown in which the isotope enrichment factors 
are strongly affected by the field shifts. 
These indications are promising towards the mass producion of enriched $^{48}$Ca 
by the chemical separation method.  
}

%
%
%
%
%

\section{Double Beta Decay of $^{48}$Ca and Chemical Separation of Calcium Isotopes}

Among $\beta\beta$ isotopes, the Q-value of $^{48}$Ca is the highest (4.27 MeV),
which is far above energies of $\gamma$-rays from natural radioactivities, therefore
our measurement is not limited by backgrounds in the energy region at 4.27 MeV.
We have developed a CaF$_2$ scintillation detector system (ELEGANT VI) 
which is an experiment with an ``active source'' (source$=$detector) containing  
$^{48}$Ca target source inside, 
being operated at the underground laboratory (Oto Cosmo Observatory) in Nara \cite{vi}.
For further improvements, we are now developing a new detector system
CANDLES(CAlcium fluoride for the study of Neutrinos and Dark matters by Low
Energy Spectrometer) sensitive to the half-life of 10$^{26}$ yr for
neutrinoless double beta decay (0$\nu\beta\beta$)
of $^{48}$Ca ($<m_{\nu}>\sim$0.03 eV), which ultimately requires
several tons of calcium \cite{candles}.
The only drawback of $^{48}$Ca in terms of 0$\nu\beta\beta$ is a small natural
abundance of 0.187 $\%$.


Now the widest variety of stable isotopes are mainly produced at electromagnetic
separators and gas centrifuges.
Practically the separation of stable isotopes of calcium is presently being accomplised
with electromagnetic separators(calutrons) at Oak Ridge National Laboratory and
Russia
\footnote{Enriched calcium is commercially available as carbonate or oxide by
TRACE Science Int., but the cost is expensive($\sim$ 200K$\$$/g)
and the amount is limited with only a few grams.}.
The rare calcium isotopes separated by this method are expensive, and their use
is often limited by cost. 
Unfortunately flexible highly efficient centrifugal technology is only possible 
for those elements
(about 20) which have gaseous compounds at room temperature.
Therefore, these methods cannot meet the production of some $\beta\beta$ isotopes
such as $^{48}$Ca, $^{96}$Zr and $^{150}$Nd etc.
The present paper shows our initial attempt toward the enrichment of $^{48}$Ca and 
its preliminary result on the separation of calcium isotope by utilizing calcium isotope
effect in liquid-liquid extraction (LLE) of calcium chloride using a crown ether.
The first publication on the chemical isotope separation with macrocyclic polyether 
was calcium isotope separation with dicyclohexano-18-crown-6(DC18C6) \cite{jep}. 
Isotopic enrichment occurs according to the following chemical exchange reaction:
\begin{equation}
^{40}Ca^{2+}_{(aq)} + ^{48}CaL^{2+}_{(org)}
\rightarrow ^{48}Ca^{2+}_{(aq)} + ^{40}CaL^{2+}_{(org)}
\end{equation}
where $L$ represents macrocyclic polyether(18-crown-6). As a result, $^{40}$Ca is
enriched in the organic-phase(org) crown solution and the heavy isotopes of $^{48}$Ca
tend to concentrate in the aqueous(aq) phase. 

\section{Experimental}

Dicyclohexano-18-crown-6 was obtained from Aldrich Chemical Company (98$\%$ purity).  
Chloroform (99$\%$ purity) and Calcium chloride (95$\%$) were products of Nacalai Tesque. 
Calcium chloride was dissolved in pure water to create a solution, 3M CaCl$_2$. 
This solution served as the aqueous phase. The organic phase was 0.07M DC18C6 in chloroform.  
A 20 ml aqueous solution and 200 ml organic solution were mixed in a flask. 
The two phases were stirred by a magnetic stirrer for 1 h and allowed 
for standing. This procedure was carried out at 280 K. 
After the two phases were separated, the upper aqueous phase was taken. 
It is noted that before the above procedure, we did a vacant-extraction by using 
a pure water to reduce impurities.  
Presently the same liquid-liquid extraction experiment was iterated six times. 

\section{Isotope Analysis}

The fractionation of $^{48}$Ca from the most abundant isotope $^{40}$Ca (96.9$\%$) is 
the key to realize the chemical separation method by liquid-liquid extraction using 
a crown ether.  
Unfortunately the most abundant $^{40}$Ca cannot be measured even by high resolution
double-focusing sector field ICP-SFMS (HR-ICP-MS; JMS-Plasmax 2) at
the Wakasa Wan Energy Research Center, where $^{42}$Ca, $^{43}$Ca and $^{48}$Ca can 
be measured \cite{oulns04}, because the coincidence 
of $^{40}$Ar$^{+}$ and $^{40}$Ca is the most annoying example of an isobaric interference 
and its required resolution(192498) is beyond this ICP-MS's maximum resolution (12000). 
In order to measure concentrations of $^{40}$Ca, reaction-cell ICP-MS(Perkin Elmer-SCIEX,  
ELAN DRCII) was used at the Center for Advanced Marine Core Research, Kochi University. 
Gas selection is an important component of interference reduction. 
The ELAN DRC system allows use of more effective reaction gases such as ammonia 
that reduces the $^{40}$Ar$^{+}$ interference for $^{40}$Ca by a factor of 
10$^9$, compared to a factor of 10$^3$ obtained by a simple ``collision'' cell ICP-MS, 
which must use simple collision gases, such as hydrogen and helium, to limit the formation 
of adverse side reaction products. 
The isotope abundance at 40, 43, 44, 48u were measured with 
anmmonia gas flow of 0.5 ml/min and nebulizer flow of 0.1 ml/min. 
The measured relative deviation from ``Cica'' 1000 ppm AAS standard 
(as a standard solution) 
of $^{40}$Ca, $^{43}$Ca, $^{44}$Ca and $^{48}$Ca are shown in Fig.~1(Left). 
The precision of measured isotopic ratios was about 0.2$\sim$0.1$\%$ (1 $\sigma$). 

The $^{40}$Ca fractionation was successfully measured, but  
the measured separation factor is a bit small, compared with the 
previous ICP-SFMS's data. 
Much more precise isotope analysis is essential for this 
confirmation and now we are on our way to verify this by using 
TIMS(Thermo Electron Corp., TRITON) 
with a double-spike technique to correct instrumental mass bias and instrumental drifts.


\section{Calcium Isotope Effect}

In the following, we will use the data from reaction-cell ICP-MS only for the consistency.
Now we can evaluate the isotope effect quantitatively by utilizing    
the new Bigeleisen theory reported in 1996 \cite{new}. 
The isotope effect in the chemical exchange reaction
can be written in terms of three effects such as the nuclear mass effect, the
nuclear size and shape effect, and the contribution of hyperfine splitting based
on the nuclear spin in the simple equation as follows, based on his new theory;

\begin{equation}
ln \alpha = a (\bigtriangleup M/MM') + b\delta<r^2> + ln K_{hf},
\end{equation}

where $M$ and $M'$ are the nuclear masses of isotopes, respectively, and
$\bigtriangleup M$ is the mass difference of isotopes, and $\delta <r^2>$
and $ln K_{hf}$ are the change in the mean square of the charge distribution
radii \cite{fricke} and the contribution of hyperfine splitting based on the nuclear spin,
respectively. The scaling factors of the nuclear mass effect and that of the
nuclear size and shape effect, namely the field shift effect, are represented
by $a$ and $b$, respectively.
By using Eq.(2), we can evaluate each contribution
of these three effects to the total calcium isotope effect.
There are three free parameters, then we need three sets of equations to get
each contribution. 
Here, our primary concern is to obtain each contribution ratios relative to 
$^{48}$Ca between the field shift effect, the mass effect, and the hyperfine 
splitting shift effect. Thus, we use the three isotope pairs of calcium,  
$^{40}$Ca$-^{48}$Ca, $^{44}$Ca$-^{48}$Ca, $^{43}$Ca$-^{48}$Ca, 

\begin{equation}
\epsilon_{40-48} = a (\bigtriangleup M/MM')_{40-48} + b\delta<r^2>_{40-48},
\end{equation}
\begin{equation}
\epsilon_{44-48} = a (\bigtriangleup M/MM')_{44-48} + b\delta<r^2>_{44-48},
\end{equation}
\begin{equation}
\epsilon_{43-48} = a (\bigtriangleup M/MM')_{43-48} + b\delta<r^2> + (ln K_{hf})_{43}.
\end{equation}

The isotope separation factor $\alpha$ and the isotope enrichment factor $\epsilon$
are defined as follows;

\begin{equation}
\alpha_{M-48} = ([^{M}Ca]/[^{48}Ca])_{org}/([^{M}Ca]/[^{48}Ca])_{aq},
\end{equation}
\begin{equation}
\epsilon_{M-48} = \alpha - 1,
\end{equation}

where ([$^{M}$Ca]/[$^{48}$Ca])$_{org}$ and ([$^{M}$Ca]/[$^{48}$Ca])$_{aq}$ are the isotopic
ratios of $^{M}$Ca to $^{48}$Ca in the organic phase and that in the aqueous phase,
respectively. The superscript $M$ means mass number 40, 44, 43, or 48.
Since the isotope enrichment factor, $\epsilon_{M-48}$ is the dimensionless,
we will use the relative values of $(\bigtriangleup M/MM')_{M-48}$ and
$\delta<r^2>_{M-48}$. 


The values of each contribution of $a$, $b$ and $ln K_{hf}$ are calculated
using sequential equations 
and compared with 
Cr$L$Cl$_3$ case, 
which was measured (and evaulated) by 
T. Fujii et al. \cite{tfujii}. 
Thus, we can estimate each contribution ratios of the field shift effect 
to the mass effect for the above three isotope pairs, which is summarized in 
Table 1, in comparison with the case of Chromium trichloride-crown(DC18C6).  
This preliminary result shows that the ratios of the field shift effect to 
the mass effect are small, especially for $^{40}$Ca$-^{48}$Ca case, compared with 
the Cr$L$Cl$_3$ case. 
This is crucial asset in order to realize the $^{48}$Ca enrichment by utilizing 
this chemical separation method, because $^{40}$Ca is the largest abundance isotope 
and if the field shift effect (the nuclear size and shape effect) is dominant, which 
is shown in filled points and lines from the nuclear charge radii in Fig. 1(Left), 
this chemical separation method is not effective for the separation of $^{48}$Ca 
from the most abundant $^{40}$Ca. 
This preliminary quantitative result on calcium isotope effect is in contrast to 
the previously obtained experimental data of chromium isotope effect \cite{tfujii}. 
Chromium isotope has a similar characteristic ``parabolic'' nuclear charge distribution 
to the calcium isotope, which are shown as filled points and lines in Fig.~1(Right). 
This is because $^{52}$Cr has the smallest nuclear charge radius due to its the same 
magic number neutrons $N=$28 as $^{48}$Ca. 
They found a clear mass-independent isotope effect that the medium $^{52}$Cr is foremost 
fractionated according to the Bigeleisen theory. 
The notable difference between Cr isotopes and Ca isotopes is that the latter has two 
doubly magic isotopes $^{40}$Ca and $^{48}$Ca and these also have a magic number of 
neutrons. 
\begin{table}[h]
\caption[]{
Contribution ratios of the field shift effect or the hyperfine 
splitting shift effect to the mass effect for $^{40}$Ca, $^{44}$Ca, and $^{43}$Ca 
relative to $^{48}$Ca, 
in comparison with that for $^{50}$Cr, $^{54}$Cr, and $^{53}$Cr 
relative to $^{52}$Cr, \cite{tfujii}. 
}
\begin{center}
\begin{tabular}{ccc} \hline
 & $b\delta<r^2>/[a (\bigtriangleup M/MM')]$ & $ln K_{hf}/[a (\bigtriangleup M/MM')]$ \\ \hline
$^{40}$Ca$-^{48}$Ca & 0.02$\pm$0.48 & - \\
$^{44}$Ca$-^{48}$Ca & 0.62$\pm$1.31 & - \\ 
$^{43}$Ca$-^{48}$Ca & 0.22$\pm$0.88 & 0.64$\pm$1.35 \\ \hline
$^{50}$Cr$-^{52}$Cr & 1.12$\pm$2.79 & - \\
$^{54}$Cr$-^{52}$Cr & $-$2.81$\pm$5.97 & - \\ 
$^{53}$Cr$-^{52}$Cr & $-$2.05$\pm$8.94 & $-$0.83$\pm$6.17 \\ \hline
\end{tabular}
\end{center}
\end{table}

\begin{figure}[hbt]
\vspace*{-1.0cm}
\vspace*{-1cm}
\leftline{\includegraphics[scale=0.35]{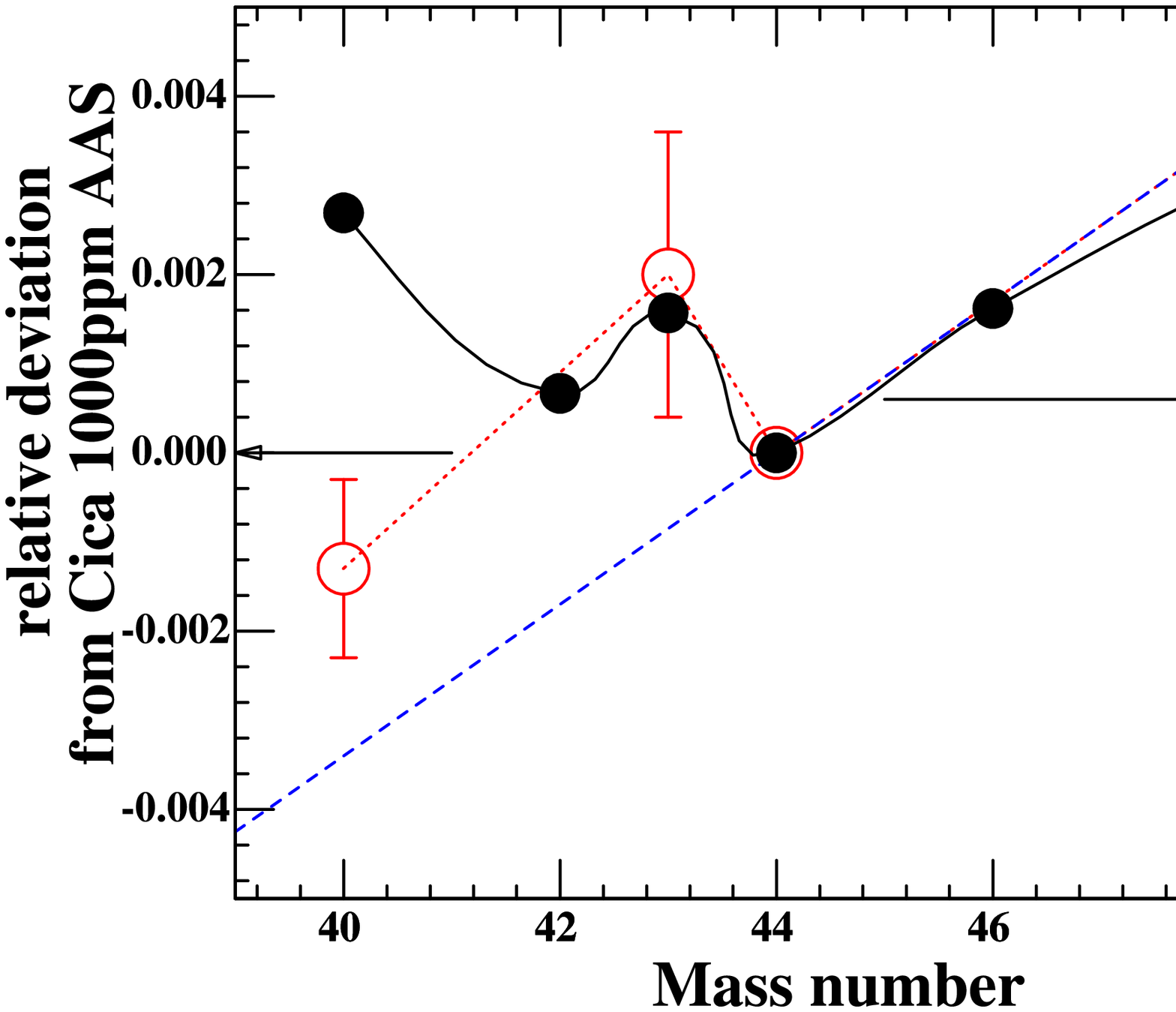}
\hspace*{3.0cm}
\includegraphics[scale=0.35]{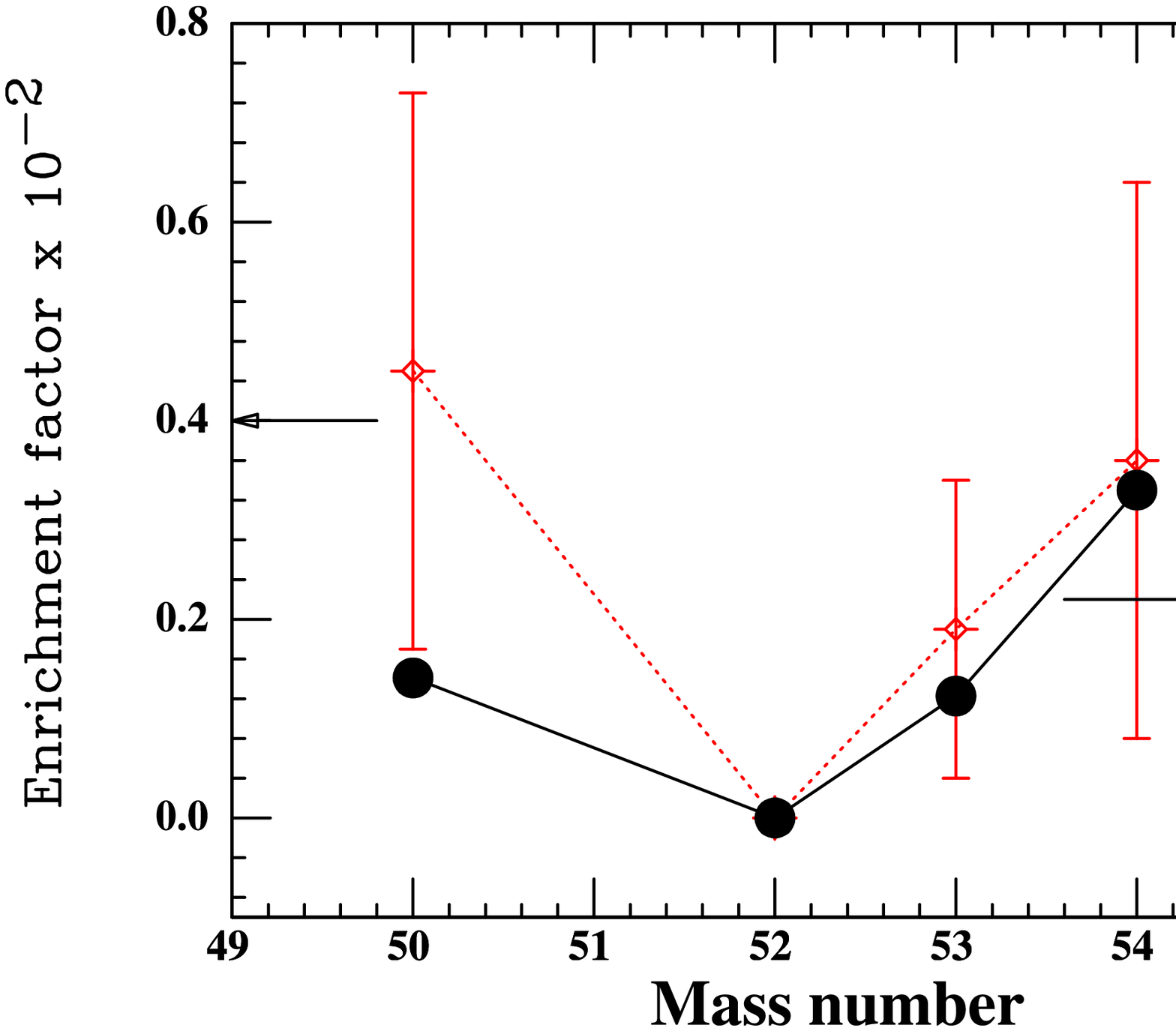}}
\vspace*{-1.5cm}
\caption{Left: Observed calcium isotope effects (left axis) and 
experimental values for the mean square nuclear charge radii \protect\cite{fricke} 
(right axis) of the calcium isotopes relative to $^{44}$Ca. 
Filled(line) and open(dotted line) points correspond to nuclear charge radii 
and our obtained data, respectively. 
The dashed line is the expected mass effect, which shows a linear relationship with mass 
number.   
Right: Previous studies on chromium isotope effects (left axis) and 
experimental values for the mean square nuclear charge radii (right axis) of the chromium 
isotopes relative to $^{52}$Cr. 
Filled(line) and square(dotted line) points correspond to nuclear charge radii 
and the experimental data measured by T. Fujii et al. \protect\cite{tfujii}, respectively. 
\label{fig:fig1}}
\end{figure}

\vspace*{-1cm} 
\section*{Acknowledgments}
This work was supported by the Ministry of Education, Science and Culture, Japan. 
The work of R.H. was supported by Furukawa Foundation (Furukawa Mfg. Co., Ltd.). 

\end{document}